\documentclass[preprintnumbers, floatfix, showkeys, preprintnumbers, letterpaper, twocolumn, superscriptaddress,nofootinbib]{revtex4-2}
\pdfoutput=1 
\usepackage{graphicx}
\usepackage{microtype}
\usepackage{amsmath}
\usepackage{amssymb}
\usepackage{subfigure}
\usepackage{url}
\usepackage{hyperref}
\usepackage{mathtools}
\usepackage{orcidlink}
\usepackage{slashbox}

\usepackage{xcolor}
\usepackage{color}
\usepackage{mathrsfs}
\usepackage{calrsfs}
\usepackage{amsfonts}
\usepackage{tabularx}
\usepackage{latexsym}
\usepackage{ragged2e}
\usepackage{epsfig}
\usepackage{textcomp}
\usepackage{float}

\usepackage{caption}
\DeclareCaptionJustification{justified}{\leftskip=0pt \rightskip=0pt \parfillskip=0pt plus 1fil}
\definecolor{vividviolet}{rgb}{0.62, 0.0, 1.0}
\definecolor{amaranth}{rgb}{0.9, 0.17, 0.31}
\definecolor{palatinateblue}{rgb}{0.15, 0.23, 0.89}
\definecolor{brightpink}{rgb}{1.0, 0.0, 0.5}
\definecolor{cornflowerblue}{rgb}{0.39, 0.58, 0.93}
\definecolor{deepcarminepink}{rgb}{0.94, 0.19, 0.22}
\definecolor{radicalred}{rgb}{1.0, 0.21, 0.37}

\hypersetup{ linktoc=all,
	colorlinks, linkcolor={palatinateblue},
	citecolor={brightpink}, urlcolor={amaranth}
}

\graphicspath{{Images/}}

\renewcommand{\d}[1]{\ensuremath{\operatorname{d}\!{#1}}}

\def\sideremark#1{\ifvmode\leavevmode\fi\vadjust{\vbox to0pt{\vss
			\hbox to 0pt{\hskip\hsize\hskip1em
				\vbox{\hsize1.3cm\tiny\raggedright\pretolerance10000
					\noindent #1\hfill}\hss}\vbox to8pt{\vfil}\vss}}}%
\def\beq{\begin{equation}}
\def\eeq{\end{equation}}

\begin{document}
\title{Corrected Hawking Temperature and Final State of Black Hole Evaporation\\ Under GEVAG Framework}

\author{Yen Chin \surname{Ong}\orcidlink{0000-0002-3944-1693}}
\email{ongyenchin@nuaa.edu.cn}
\affiliation{Center for the Cross-disciplinary Research of Space Science and Quantum-technologies (CROSS-Q), College of Physics, Nanjing University of Aeronautics and Astronautics, \\29 Jiangjun Road, Nanjing City, Jiangsu Province 211106, China}

\begin{abstract}
In the GEVAG (Generalized Entropy Varying-G) framework, any generalization to horizon entropy leads to a varying gravitational ``constant'' $G_\text{eff}$ that is a function the horizon area.
In this work, it is shown that if we promote $G_\text{eff}$ to be valid in the neighborhood of the horizon, then Hawking temperature consists of two terms, the second of which is related to the variation of $G_\text{eff}$. When applied to the logarithmic correction of the entropy, as is common across various quantum gravity approaches,
the first term in the Schwarzschild black hole temperature exactly agrees with that obtained from utilizing generalized uncertainty principle (GUP), while the second term improves on the GUP result by driving the Hawking temperature to zero as the black hole approaches a minimum mass. This resolves the inconsistency in the GUP result concerning a nonzero temperature minimum mass remnant. This work also derives simple general formulas for both the thermodynamic energy and the Bekenstein bound for any correction to the area law under the same assumption that $G_\text{eff}$ can be extended off-shell in the horizon neighborhood. The (generalized) Bekenstein bound can be interpreted as a statement regarding the renormalization group scaling dimension of the entropy functional $f(A)$ and the naturalness of the theory.
\end{abstract}

\maketitle

\section{Introduction: GEVAG and GUP}

Following Jacobson's method of deriving Einstein's field equations \cite{9504004}, an approach called GEVAG (Generalized Entropy Varying-G) was introduced in \cite{2407.00484}. It is found that if the Bekenstein-Hawking area law for horizon is generalized from $S=A/4G$ to $S=f(A)/4G$, then the consistent theory of gravity has field equation of the form (if cosmological constant is zero)\footnote{We shall work with the units that $\hbar=c=1$ but $G$ is left explicit.},
\begin{equation}
R_{ab}-\frac{1}{2}g_{ab}R = {8\pi G_\text{eff}} T_{ab}.
\end{equation}
The only difference from standard general relativity is the gravitational constant $G$ has been replaced by an effective one, related to the original $G$ by
\begin{equation}\label{Geff}
G_\text{eff} = \frac{G}{f'(A)},
\end{equation}
where $f'(A)$ denotes the derivative with respect to the horizon area. Of course, a sensible theory should have $f'(A) >0$.
Crucially $G_\text{eff}$ becomes area-dependent, i.e., this is a type of varying-$G$ gravity theory. Alternatively one may view it as GR but with a non-trivial matter-coupling.
For the various subtleties of how one might interpret the physics in such a theory (for example, how this can be consistent with observations, what happens when there is no horizon, or when there are multiple horizons), see \cite{2407.00484}. See also the more current work \cite{2602.20430} in which the author linked $G_\text{eff}(A)$ with the idea of ``topological calibration''.

In \cite{2505.07972}, it was shown that if we considered a logarithmic correction to the entropy (which arises in many approaches to quantum gravity)
\begin{equation}
f(A)=A+c_1 \ln\left(\frac{A}{G}\right),
\end{equation}
the GEVAG frameworks yields, via Eq.(\ref{Geff}),
\begin{equation}\label{Geff2}
G_\text{eff}=\frac{G}{1+\frac{c_1}{A}}.
\end{equation}
By utilizing the horizon area $A=16\pi G_\text{eff}^2 M^2$, we could obtain an explicit form of $G_\text{eff}$ as given in \cite{2505.07972}:
\begin{equation}
G_\text{eff} = \frac{1}{4}\left(\frac{2\pi GM + \sqrt{4\pi^2 G^2M^2-\pi c_1}}{\pi M}\right).
\end{equation}
One then obtains the modified horizon radius
\begin{equation}\label{hor}
r_h = 2G_\text{eff}M =GM + \sqrt{G^2M^2-\frac{\alpha G}{4}}.
\end{equation} 
The Hawking temperature was taken to be $T=1/8\pi G_\text{eff}M$ (i.e. the same form as in GR except $G$ is replaced by $G_\text{eff}$) in \cite{2505.07972}, which gives 
\begin{equation}
T=\frac{1}{8\pi G_\text{eff} M} = \frac{1}{4\pi GM + 2\sqrt{4\pi^2G^2M^2-c_1\pi}}.
\end{equation}
This is algebraically equivalent to
\begin{equation}\label{TGUP}
T=\frac{M}{\pi \alpha}\left(1-\sqrt{1-\frac{\alpha}{4GM^2}}\right), 
\end{equation}
the Hawking temperature one obtains by heuristic argument from the generalized uncertainty principle \cite{0106080}, if we identify $c_1=\pi G \alpha$, where $\alpha$ is the GUP parameter in 
\begin{equation}\label{GUP}
\Delta x \Delta p \geqslant \frac{1}{2}\left(\hbar + \alpha l_\text{Pl}^2 \frac{\Delta p^2}{\hbar}\right),
\end{equation} 
where $l_\text{Pl}$ is the Planck length (we restored $\hbar$ here for clarity).
This is itself already quite remarkable, since just by using the logarithmic correction, the GEVAG framework reproduces the GUP result without explicitly using any GUP. 

In this work, we shall see that if we extend $G_\text{eff}(A)$ to be valid not only for horizon, but also for $(r-\varepsilon, r+\varepsilon)$ in the vicinity of the horizon, then 
GEVAG actually \emph{improves} upon the GUP result, not merely reproducing it. This is due to the fact that
the full Hawking temperature is then \emph{not} just $T=1/8\pi G_\text{eff}M$ as used in Ref.\cite{2505.07972}, but contains an extra term due to the ``running'' of $G_\text{eff}$. This allows the Hawking temperature to go to zero as the black hole approaches the minimum mass. This in turn solves a problem faced by the GUP expression\footnote{In addition, at least in the original GUP black hole case \cite{0106080}, the horizon radius was unmodified, so the minimum mass was not reflected by the metric. In our case, the minimum mass follows from both the temperature and horizon expression, Eq.(\ref{hor}). }: in order to ensure $T \in \Bbb{R}$ in Eq.(\ref{TGUP}), the black hole must stop evaporating at the minimum mass $M_\text{min}=\sqrt{\alpha/4G}$, but this is inconsistent with the $T \neq 0$ state\footnote{Nonzero temperature remnant also occurred in Einstein-dilaton-Gauss-Bonnet gravity. Again it was pointed out that this is puzzling and the true final end state is a nontrivial issue \cite{2205.13006}.}. Let us now examine how GEVAG solves this problem.

\section{Full Temperature Under GEVAG}

Since GEVAG stemmed from the Jacobson's method in its derivation, it has been assumed that the Hawking temperature is given by the surface gravity via $T=\kappa/2\pi=[g'(r_h)/2]/2\pi$, where $g(r)=-g_{tt}$ is the metric function and $r_h=2 G_\text{eff}M$. (Whether prime denotes derivative with respect to $r$ or $A$ will be clear from the context). In GEVAG, $G_\text{eff}$ is only defined on the horizon, thus when differentiating terms like $G_\text{eff}M/r$ with respect to $r$, the implicit assumption in \cite{2407.00484, 2505.07972} is to treat $G_\text{eff}$ as a constant. This is like treating $G_\text{eff}$ as a state parameter of the thermodynamics. On the other hand, when a geometry is highly dynamical, especially towards the end of Hawking evaporation, perhaps $G_\text{eff}$ should be varying, and thus promoting it to 
\begin{equation}
G_\text{eff} = \frac{G}{f(4\pi r^2)}, 
\end{equation}
where $r$ is not necessarily the horizon is physically motivated. The problem is that the theory does not provide any dynamics for $G_\text{eff}$ defined away from a horizon. Such an off-shell extension would need to be taken as an extra assumption and could in fact cause more problems vis-à-vis how to interpret the effective gravitational ``constant'' from the point of view of observations, discussed in \cite{2407.00484}. Let us therefore take a middle ground by only consider this extension in the neighborhood of the horizon and \emph{not} at arbitrary $r$. This is sufficient for the following calculations. In fact, it is not uncommon to study horizon physics by varying quantities initially defined on the horizon within a small neighborhood (e.g. the ``stretched horizon'').

Now that we have justified extending $G_\text{eff}$ to $G_\text{eff}(r)$ in the horizon neighborhood, 
since $G_\text{eff}$ is varying we should have two terms once the derivative is taken. Explicitly,
This allows us to write
\begin{equation}
g'(r_h) = \frac{1}{2G_\text{eff}M}  - \frac{1}{G_\text{eff}} \left.\frac{\d G_\text{eff}}{\d r}\right\vert_{r=r_h}.
\end{equation}
Consequently, the Hawking temperature picks up a second term involving the derivative of $G_\text{eff}$:
\begin{equation}
T=\frac{1}{4\pi}\left(\frac{1}{r_h}-\left.\frac{G'_\text{eff}}{G_\text{eff}}\right\vert_{r=r_h}\right) =\frac{1}{8\pi G_\text{eff}M}  - \frac{1}{4\pi} \left.\frac{G'_\text{eff}}{G_\text{eff}}\right\vert_{r=r_h}.
\end{equation}
As shown in \cite{2505.07972}, the first term is identical to the GUP expression, so we call it $T_\text{GUP}$. The second term shall be referred to as the correction term, denoted by $+T_\text{correction}$. That is, the full Hawking temperature is
\begin{equation}
T= T_\text{GUP}+T_\text{correction}.
\end{equation}
Explicitly, $T_\text{GUP}$ at the minimum mass can be evaluated from Eq.(\ref{TGUP}) and is given by
\begin{equation}
T_\text{GUP}=\frac{1}{2\sqrt{\pi c_1}}.
\end{equation}

For our case, the effective gravitational coupling is
\begin{equation}
G_\text{eff} = \frac{G}{1+\frac{c_1}{4\pi r_h^2}},
\end{equation}
from which we can obtain
\begin{equation}
G'_\text{eff} = \frac{G}{u^2}\frac{c_1}{2\pi r_h^3}, 
\end{equation}
where we have denoted the denominator of $G_\text{eff}$ as $u(r_h):=1+{c_1}/{4\pi r_h^2}$.
This implies the corrected term of the temperature should be
\begin{equation}
T_\text{correction}= -\frac{1}{4\pi G_\text{eff}}\left(\frac{G}{u^2}\frac{c_1}{2\pi r_h^3}\right)=-\frac{1}{4\pi u}\frac{c_1}{2\pi r_h^3}.
\end{equation}

At the minimum mass, recall again that $M_\text{min}=\frac{1}{2G}\sqrt{\frac{c_1}{\pi}}$, and furthermore the minimum area is 
\begin{equation}
A_\text{min}= 4\pi (GM_\text{min})^2 = c_1. 
\end{equation}
Thus $u = 1+c_1/A = 1+1=2$. This gives
\begin{equation}
T_\text{correction}= -\frac{1}{4\pi \cdot 2}\frac{c_1}{2\pi (GM_\text{min})^3}=-\frac{1}{2\sqrt{\pi c}}.
\end{equation}

We note the remarkable cancellation in this limit:
\begin{equation}
T= T_\text{GUP}+T_\text{correction} = \frac{1}{2\sqrt{\pi c}}-\frac{1}{2\sqrt{\pi c}}=0.
\end{equation}
Thus, unlike the GUP case for which the temperature is finite at a supposedly fixed mass remnant, we have an effective remnant that slowly evaporates asymptotically towards the zero temperature state. 

\begin{figure}[!h]
\centering
\includegraphics[width=0.50\textwidth]{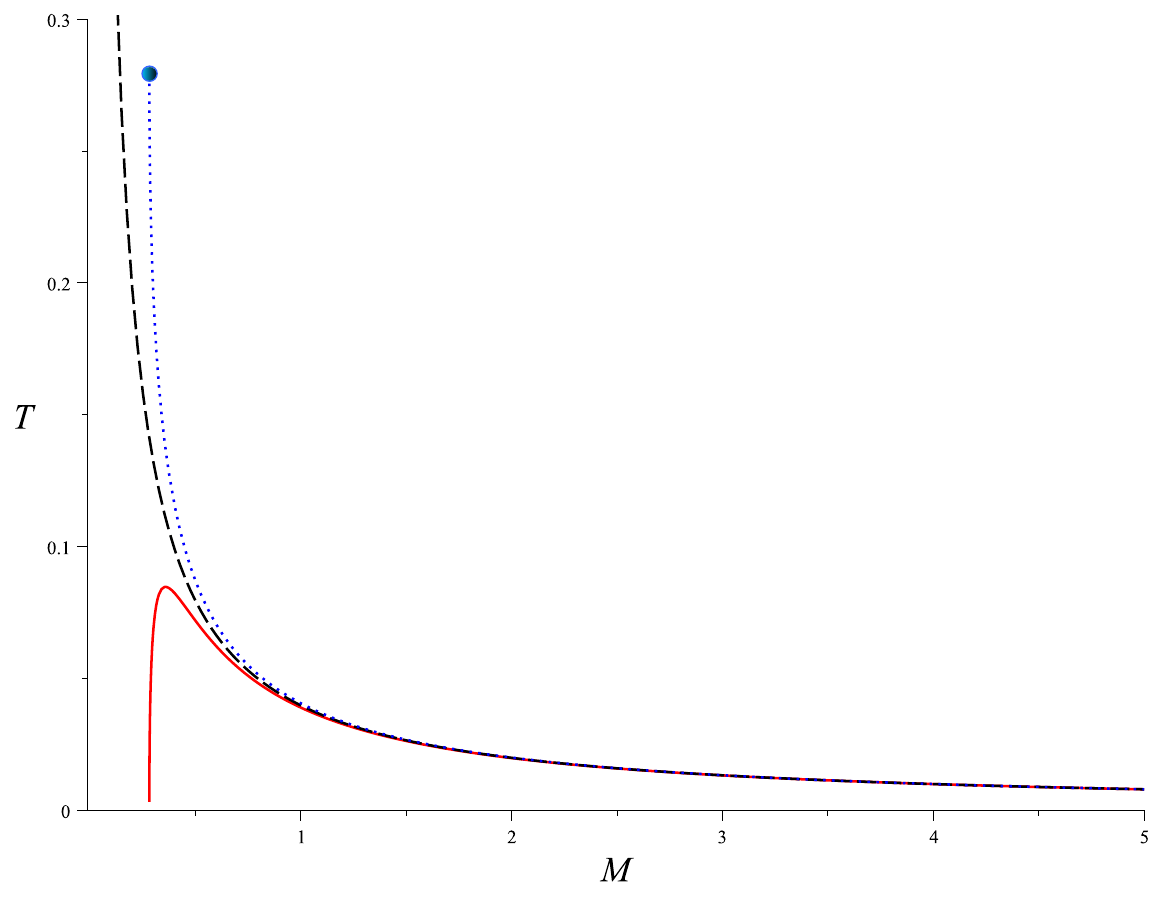}
\caption{The Hawking temperature (as a function of $M$) for the fully corrected Hawking temperature under GEVAG, assuming logarithmic correction to the Bekenstein Hawking entropy shown in sold red curve. It goes to zero as $M \to M_\text{min}$. This is compared to the dashed black curve that shows the original Hawking expression which is inversely proportional to the mass, as well as the dotted blue curve for the GUP expression that ends with a finite temperature remnant indicated by the blue dot. In this plot we set $G=c_1=1$ for numerical convenience. \label{plot}}
\end{figure}

We can plot the graph of the full Hawking temperature and compare them with the original Hawking expression as well as the GUP expression. This is shown in Fig.(\ref{plot}).

We can now compute the thermodynamic energy $E$ via the first law $\d E = T\d S$. This is found to be
\begin{equation}
E(A) = \frac{\sqrt{A}}{4G\sqrt{\pi}}\left(1+\frac{c_1}{A}\right) = \frac{r_h}{2G} \left(1+\frac{c_1}{A}\right).
\end{equation}
We note that in GR, since $c_1 = 0$, we re-cover the equivalence between ADM mass and thermodynamic energy: $E=M$.

\begin{figure}[!h]
\centering
\includegraphics[width=0.50\textwidth]{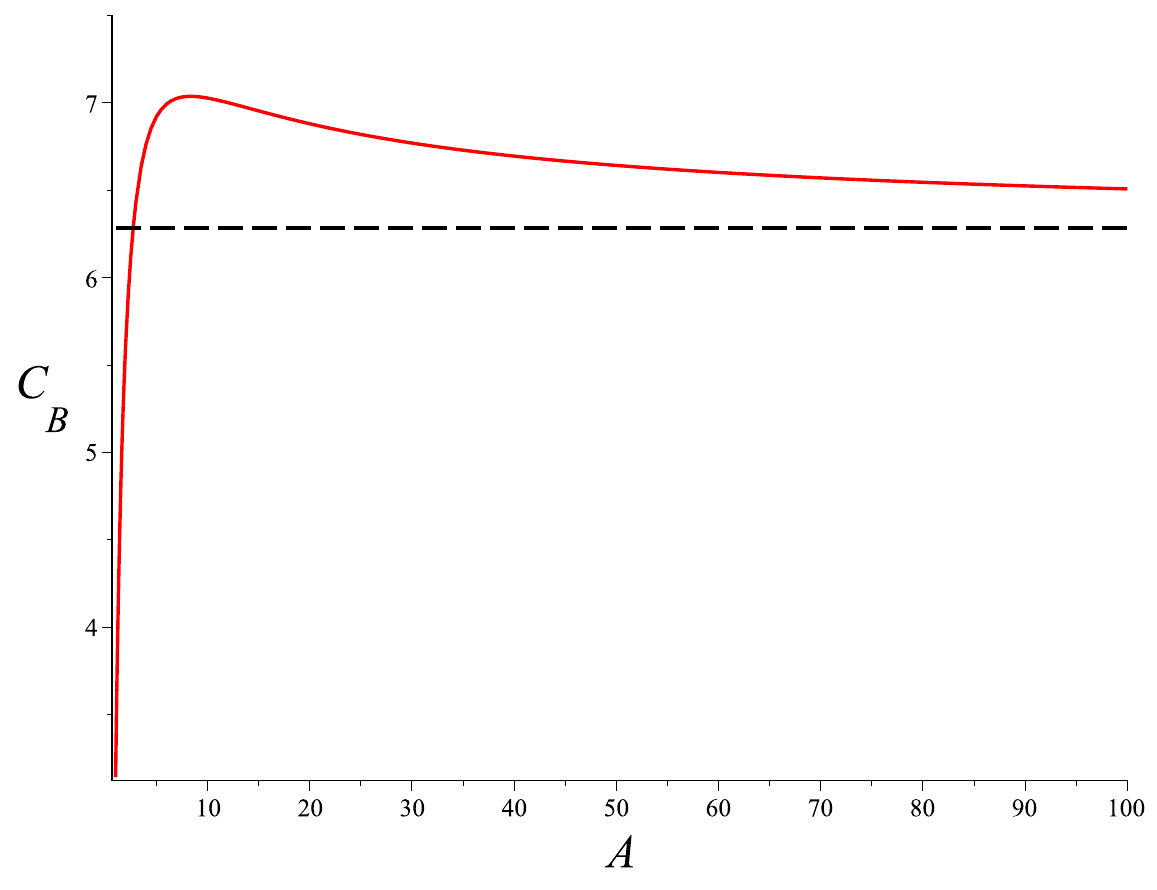}
\caption{The Bekenstein constant as a function of the horizon area. Here we take $c_1=1$. In the large $A$ limit it asymptotes from above to the GR value of $2\pi$. When $A=1$, the minimum area, we have $C_B=\pi$. \label{CB}}
\end{figure}

Finally, let us note that since the previous works only work with $T=1/8\pi G_\text{eff} M$ and not the full temperature expression, the discussions concerning the validity of the Bekenstein bound \cite{2505.07972} needs to be re-evaluated. The Bekenstein constant $C_B$, in the equation $S \leqslant C_B RE$, where $R\equiv r_h$ is the horizon (black holes saturate this inequality), is found to be simpler than the one given in  \cite{2505.07972}. We now express $C_B$ in terms of the horizon area:
\begin{equation}
C_B (A) = 2\pi \left(\frac{A+c_1 \ln A}{A+c_1}\right).
\end{equation}
This function has a global maximum that can be expressed in terms of the Lambert-W function. In the large $A$ limit the GR result is recovered $C_B \to 2\pi$. In quantum gravity, $c_1$ is often of order unity, for which the deviation of $C_B$ from $2\pi$ is small. In Fig.(\ref{CB}) we take $c_1=1$ and note that the largest deviation occurs at the minimum area, for which $C_B=\pi$.

\section{General Formulas for Arbitrary Entropy Correction}

Given any entropy correction $S=f(A)/4G$,
the full Hawking temperature can be stated as
\begin{equation}
T = \frac{f'(A)}{8\pi G M}+ \frac{4GM f''(A)}{f'(A)^2},
\end{equation}
in which prime denotes derivative with respect to $A$. 
Equivalently\footnote{For our log-corrected entropy, the temperature in terms of area takes a simple form: $T(A)=\frac{1}{\sqrt{\pi A}}\frac{A-c_1}{2(A+c_1)}$.}, 
\begin{equation}
T = \frac{1}{4\pi r_h}+ \frac{2r_h f''(A)}{f'(A)}=\frac{1}{2\sqrt{\pi A}}+\sqrt{\frac{A}{\pi}}\frac{f''(A)}{f'(A)}. ~~ 
\end{equation}
Thus given any correction to the area law, we can directly compute its Hawking temperature.

For other generalized entropies as investigated in \cite{2505.03907}, the analysis concerning Bekenstein bound therein should also be re-evaluated quantitatively if we extend $G_\text{eff}$ off-shell. Still, qualitatively we do not expect violation of the spirit of Bekenstein bound in the sense that $C_B$ is always bounded. This can be argued as follows: the energy expression for general given entropy $S=f(A)/4G$ is simply given by (but recall that $A$ contains $G_\text{eff}$, which itself is defined in terms of $A$, solving the explicit form can be algebraically tedious)
\begin{equation}\label{EA}
E(A) = \frac{\sqrt{A}f'(A)}{4\sqrt{\pi}G}.
\end{equation}
To see this, we note that from the first law $\d E=T\d S$, we can obtain
\begin{equation}
\d E=\frac{1}{4\sqrt{\pi}G} \left[\frac{f'(A)}{2\sqrt{A}}+\sqrt{A}f''(A)\right]\d A.
\end{equation}
Further note that the square bracket term is a total differential $\d(\sqrt{A}f'(A))/\d A$. If we choose the integration constant to be zero for $A \to 0$, this gives Eq.(\ref{EA}) above.
In GR, $f'(A)=1$ and Eq.(\ref{EA}) reduces to $E = M$.

Consequently, we can compute the the Bekenstein constant in terms of the area as
\begin{equation}\label{E}
C_B(A) = \frac{S(A)}{R(A)E(A)}=2\pi \frac{f(A)}{Af'(A)}.
\end{equation}
This is always finite if the entropy function $f(A)$ is well-defined, and if $f'(A) \neq 0$, i.e. $G_\text{eff}$ does not diverge. Such a divergence can actually happen, for example in the logarithmically corrected entropy case for $c_1 <0$, since $G_\text{eff} = G/(1-|c_1|/A)$ diverges at $A=|c_1|$. If we insist that $C_B$ should be $O(1)$ multiple of $2\pi$ (the weak form of Bekenstein bound, as categorized in \cite{2505.03907}), we can constrain the parameters in a given generalized entropy.

In fact, $C(B)/2\pi \sim O(1)$ means that $f(A)/Af'(A) \sim O(1)$. We can define the inverse ratio\footnote{The notation $\gamma$ is suggestive; it follows that of the usual notation for anomalous dimension in quantum field theories.}
\begin{equation}
\gamma(A) := \frac{Af'(A)}{f(A)} = \frac{\d(\ln f(A))}{\d(\ln A)}.
\end{equation}
This can be loosely interpreted as a RG-scaling dimension of the entropy functional $f(A)$. Thus interestingly,
\emph{the (generalized) Bekenstein bound can therefore be interpreted as a statement regarding the renormalization group (RG)-scaling dimension and the naturalness of the theory} (given a specific form of $f(A)$). Namely, if $\gamma(A) \sim 1$, we have a natural scaling for the entropy and there is no large or small parameters hidden in $f$ or fine-tuned hierarchy. Note that GR satisfies $\gamma=1$. For any power law like correction $f(A)=A^p$ (such as the Tsallis-Citro entropy \cite{1202.2154}), we get $\gamma=p$. Weak form of Bekenstein bound requires $p\sim O(1)$.

The $f'(A)=0$ case, in terms of the RG-language, would be a critical point with singular RG-behavior. It would be interesting to study the behavior of GEVAG in terms of a more precise RG-language in future works and so compare it with the traditional approach of asymptotically safe gravity (ASG). 

It is worth mentioning that the link between GEVAG and ASG was already noted in \cite{2407.00484}. Specifically, the form of $G_\text{eff}$ in Eq.(\ref{Geff2}) does indeed take the form of running-$G$ in ASG:
\begin{equation}
G(k)=\frac{G(k_0)}{1+Ck^2},
\end{equation}
for some constant $C$, if the energy scale $k \propto 1/\sqrt{A}$, precisely what was argued for in \cite{2204.09892}. In addition, possible connections between GUP and ASG were also investigated in \cite{2204.07416}.

\section{Conclusion: GEVAG Improves GUP}

From the above calculation, we see that if we promote $G_\text{eff}$ to be valid off-shell, at least in the $\varepsilon$-neighborhood of the horizon, 
taking into account GEVAG consistently gives rise to two terms in the Hawking temperature, the first of which matches exactly the GUP expression as found in \cite{2505.03907}. However, it also picks up a second term involving the derivative of $G_\text{eff}$. This term is responsible for driving the Hawking temperature to zero in the final stages of the evaporation. Without this term, the evaporation according to GUP stops at finite temperature, which is peculiar considering that $M$ is supposed to stay fix at $M_\text{min}$. In \cite{2505.03907}, I already argued that this GUP behavior may indicate that it is still not fully consistent and one may need to consider higher order term correction to the entropy beyond the logarithmic one. However, this work shows that this is not necessary. In fact, just by making a reasonable assumption on $G_\text{eff}$ near horizon, the final picture is modified to a sensible one: the temperature turns around once reaching a maximum and eventually reduces to zero. This is the picture that is consistent with many different approaches in the literature, including but not limited to non-commutative geometry inspired models \cite{0510112}, some asymptotically safety models \cite{0602159,1401.4452}, the ``quantum Oppenheimer-Snyder'' model \cite{2859197}, effective loop quantum black holes in the polymer models \cite{2310.01560, 2504.06964}, string theoretic models \cite{2203.10957,1902.11242}, gravity's rainbow models \cite{1402.5320v2}, quantum Raychaudhuri equation approach in Bohmian mechanics \cite{1509.02495} and black holes in Finsler geometry \cite{s10052}. Hawking evaporation in the context of varying-G theories (but not GEVAG) has also been studied. For a recent example, see \cite{2601.17162}.

Given that an extra assumption could a priori cause the temperature to be modified in any way, the fact that it causes an exact cancellation to obtain zero temperature at the minimum mass in turn supports our assumption about the off-shell validity of $G_\text{eff}$ in the vicinity of the horizon. 

To conclude, GEVAG shows explicitly that GUP temperature originated from changing $G$ to $G_\text{eff}$ in the standard Hawking formula $1/(8\pi G_\text{eff} M)$, and then also shows how its pathology is cured from the second correction term involving $G'_\text{eff}$.

On the other hand, we should mention that the above nice result is only for $c_1 > 0$. For the opposite sign, which corresponds to GUP with negative parameter $\alpha$, the full Hawking temperature now diverges because
\begin{equation}
T_\text{correction} = -\frac{|c_1|}{8\pi^2 r_h^3 u},
\end{equation}
but the denominator
\begin{equation}
u=1-\frac{|c_1|}{4\pi r_h^2} \to 0
\end{equation}
when $r \to r_\text{min}$.
Note that in this case the horizon is located at
\begin{equation}
r_h = GM + \sqrt{G^2M^2 + \frac{|c_1|}{4\pi}}, 
\end{equation}
so in the $M \to 0$ limit\footnote{Note that there is no minimum mass here despite the final area is nonzero, c.f. the result in \cite{1806.03691}, in which the effect of negative GUP parameter on evaporating Schwarzschild black hole was investigated. In that approach, just like the positive parameter case, there is no $T_\text{correction}$, only $T_\text{GUP}$. It was found that the zero mass limit has a finite nonzero temperature, but the evaporation time is infinite. I argued therein that this means such a pathological state would therefore be unattainable. However, this does not explain why such zero mass black holes cannot be copiously produced by quantum fluctuations. GEVAG provides an explanation: the negative GUP parameter case is not physical once $G'_\text{eff}/G_\text{eff}$ term is included. This is further supported by the divergence in the Bekenstein constant in this limit as we have mentioned above.}, we get $r \to r_\text{min}=\sqrt{|c_1|/4\pi}$.
At the same time, as noted in the last section, the Bekenstein bound is no longer satisfied even in the weak form as $G_\text{eff} \to \infty$ in this limit. 

From this perspective, it seems that the $c_1 < 0$ case, although was seriously considered in the literature and in fact favored from some other considerations (see e.g. \cite{1804.05176,0912.2253,1903.01382v1,2511.15871,grg,2603.23551}), is less favorable unless the situation changes when higher order terms are included. Another possibility is that $G_\text{eff}$ does not truly diverge: since Hawking emission is discrete, the black hole may evolve from $G_\text{eff}>0$ to $G_\text{eff}<0$ via a ``jump''. With the flip in the sign of $G_\text{eff}$, it could be possible to render Hawking temperature negative. Such a scenario was noticed in the string T-duality black hole model \cite{1902.11242}, in which the sign of the temperature oscillates. While black holes could not reach this state if emission is smooth (as they would end at the $T=0$ remnant state), discrete emission may make this transition possible. Of course negative temperature is not physical; this could signal some kind of instability or phase transition as the end state, instead of an effective remnant as in the $c_1 > 0$ case. Perhaps one could also entertain the possibility that $c_1$ and therefore $\alpha$ is not fixed and can change sign depends on energy scale or field content, but this is beyond the scope of the current work. Such a possibility was hinted at in \cite{1408.3763}. See also \cite{1801.03670,2870478} in which the GUP parameter depends on the system. 

Note that we are \emph{not} claiming that logarithmic correction necessarily means that we should vary the gravitational constant, as this is certainly not done in established approaches like string theory or loop quantum gravity. GEVAG is meant as an \emph{effective} classical theory of gravity. When logarithmic corrected entropy is used in GEVAG, it tells us what the effective metric should be (thus also avoids the unreliable heuristic approaches in the GUP literature to obtain an effective metric \cite{2303.10719}), and this requires $G$ to be varied. In the full QG, this is not necessarily the case.

\end{document}